\documentclass[preprint,12pt,authoryear]{elsarticle}

\usepackage{mathabx}
\usepackage{hyperref}
\usepackage{color}
\usepackage{soul}

\newcommand\aj{{AJ}}
\newcommand\apj{{ApJ}}
\newcommand\apjl{{ApJL}}     
\newcommand\apjs{{ApJS}}
\newcommand\mnras{{MNRAS}}
\newcommand\pasj{{PASJ}}
\newcommand\sovast{{Soviet~Ast.}}
\newcommand\pasa{{PASA}}

\newcommand{\RA}{\textit{RadioAstron}\ }

\journal{Advances in Space Research}
\usepackage[ps2pdf,%
a4paper=true,%
breaklinks=true,%
colorlinks=true,%
pdfauthor={Kovalev et al.},%
pdftitle={RadioAstron AGN survey}%
]{}

\begin{document}
\begin{frontmatter}
\title{Detection statistics of the RadioAstron AGN survey}
\author[a1,a2,mpi]{Y.~Y.~Kovalev}
\author[a1]{N.~S.~Kardashev\footnote{Deceased}} 
\author[a1,a4,a5]{K.~V.~Sokolovsky} 
\author[a1]{P.~A.~Voitsik} 
\author[shao]{T.~An} 
\author[g1,g2]{J.~M.~Anderson} 
\author[a1]{A.~S.~Andrianov} 
\author[a1]{V.~Yu.~Avdeev} 
\author[york]{N.~Bartel} 
\author[au3]{H.~E.~Bignall} 
\author[a1]{M.~S.~Burgin} 
\author[au2]{P.~G.~Edwards} 
\author[au4]{S.~P.~Ellingsen} 
\author[h1]{S.~Frey} 
\author[ska]{C.~Garc\'ia-Mir\'o} 
\author[torun]{M.~P.~Gawro\'nski} 
\author[gb]{F. D. Ghigo} 
\author[gb,arecibo]{T.~Ghosh} 
\author[i1,i2]{G.~Giovannini} 
\author[a1]{I.~A.~Girin} 
\author[i1]{M.~Giroletti} 
\author[jive,delft]{L.~I.~Gurvits} 
\author[au2,au5]{D.~L.~Jauncey} 
\author[au1]{S.~Horiuchi} 
\author[iaa]{D.~V.~Ivanov} 
\author[iaa]{M.~A.~Kharinov} 
\author[t1]{J.~Y.~Koay} 
\author[a1]{V.~I.~Kostenko} 
\author[a3]{A.~V.~Kovalenko} 
\author[a1]{Yu.~A.~Kovalev} 
\author[i1,a1]{E.~V.~Kravchenko} 
\author[torun]{M.~Kunert-Bajraszewska} 
\author[a1,astron]{A.~M.~Kutkin} 
\author[a1]{S.~F.~Likhachev} 
\author[mpi,a1]{M.~M.~Lisakov} 
\author[a1]{I.~D.~Litovchenko} 
\author[au4]{J.~N.~McCallum} 
\author[i3]{A.~Melis} 
\author[iaa]{A.~E.~Melnikov} 
\author[i3]{C.~Migoni} 
\author[jive]{D.~G.~Nair} 
\author[a1]{I.~N.~Pashchenko} 
\author[au2]{C.~J.~Phillips} 
\author[astron]{A.~Polatidis} 
\author[a1,crao]{A.~B.~Pushkarev} 
\author[hart]{J.~F.~H.~Quick} 
\author[iaa]{I.~A.~Rakhimov} 
\author[au3]{C.~Reynolds} 
\author[robledo]{J.~R.~Rizzo} 
\author[a1]{A.~G.~Rudnitskiy} 
\author[f1,f2,mpi]{T.~Savolainen} 
\author[a1]{N.~N.~Shakhvorostova} 
\author[a1]{M.~V.~Shatskaya} 
\author[shao,ch2]{Z.-Q.~Shen} 
\author[a1]{M.~A.~Shchurov} 
\author[astron]{R.~C.~Vermeulen} 
\author[s1]{P. de Vicente} 
\author[torun]{P.~Wolak} 
\author[mpi]{J.~A.~Zensus} 
\author[a1]{V.~A.~Zuga} 

\address[a1]{Astro Space Center of Lebedev Physical Institute, Profsoyuznaya~St.~84/32, 117997~Moscow, Russia}
\address[a2]{Moscow Institute of Physics and Technology, Dolgoprudny, Institutsky per., 9, Moscow region, 141700, Russia}
\address[mpi]{Max-Planck-Institut f\"ur Radioastronomie, Auf dem H\"ugel 69, 53121 Bonn, Germany}
\address[a4]{Department of Physics and Astronomy, Michigan State University, East Lansing, Michigan 48824, USA}
\address[a5]{Sternberg Astronomical Institute, Moscow State University, Universitetskii~pr.~13, 119992~Moscow, Russia}
\address[shao]{Shanghai Astronomical Observatory, Chinese Academy of Sciences, Shanghai 200030, Peopleʼs Republic of China}
\address[g1]{Institute of Geodesy and Geoinformation Science, Technical University of Berlin, Straße des 17. Juni 135, 10623 Berlin,
Germany}
\address[g2]{Department of Geodesy, GFZ German Research Centre for Geosciences, Telegrafenberg, 14473 Potsdam, Germany}
\address[york]{York University, 4700 Keele Street, Toronto, ON, M3J 1P3, Canada}
\address[au3]{CSIRO Astronomy and Space Science, PO Box 1130, Bentley WA 6102, Australia}
\address[au2]{CSIRO Astronomy and Space Science, PO Box 76, Epping, NSW 1710, Australia}
\address[au4]{School of Natural Sciences, Private Bag 37, University of Tasmania, Hobart 7001, TAS, Australia}
\address[h1]{Konkoly Observatory, MTA Research Centre for Astronomy and Earth Sciences, Konkoly Thege M. \'ut 15-17, 1121 Budapest, Hungary}
\address[ska]{Square Kilometre Array Organisation (SKAO), Jodrell Bank Observatory, Lower Withington, Macclesfield, Cheshire SK11 9DL, United Kingdom}
\address[torun]{Centre for Astronomy, Faculty of Physics, Astronomy and Informatics, NCU, Grudziacka 5, 87-100 Toru\'n, Poland}
\address[gb]{Green Bank Observatory, P.O. Box 2, Green Bank, WV~24944, USA}
\address[arecibo]{Arecibo Observatory, HC03 Box 53995, Arecibo PR 00612}
\address[i1]{INAF Istituto di Radioastronomia, via Gobetti 101, I-40129 Bologna, Italy}
\address[i2]{Dipartimento di Fisica e Astronomia, Universit\`a di Bologna, via Gobetti\ 93/2 40129 Bologna, Italy}
\address[jive]{Joint Institute for VLBI ERIC, Oude Hoogeveensedijk 4, 7991 PD,
Dwingekoo, The Netherlands}
\address[delft]{Department of Astrodynamics and Space Missions, Delft University of Technology, Kluyverweg 1, 2629 HS Delft, The Netherlands}
\address[au5]{Research School of Astronomy \& Astrophysics, Australian National University, Canberra ACT, Australia}
\address[au1]{CSIRO Astronomy and Space Science, Canberra Deep Space Communication Complex\ PO Box 1035, Tuggeranong, ACT 2901, Australia}
\address[iaa]{Institute of Applied Astronomy, Russian Academy of Sciences, nab.~Kutuzova\ 10, 191187 St.~Petersburg, Russia}
\address[t1]{Institute of Astronomy and Astrophysics, Academia Sinica, PO Box 23-141, Taipei 10617, Taiwan}
\address[astron]{ASTRON, Netherlands Institute for Radio Astronomy, Oude Hoogeveensedijk 4, 7991~PD Dwingeloo, the Netherlands}
\address[a3]{Pushchino Radio Astronomy Observatory of Astro Space Center of Lebedev Physical Institute, Moscow region, 142290 Pushchino, Russia}
\address[i3]{Cagliari Astronomical Observatory of National Institute for Astrophysics, Viale della scienza \ 5, 09047 Selargius, Italy}
\address[ch2]{Key Laboratory of Radio Astronomy, Chinese Academy of Sciences, Nanjing 210008, Peopleʼs Republic of China}
\address[crao]{Crimean Astrophysical Observatory, Nauchny 298409, Russia}
\address[hart]{Hartebeesthoek Radio Astronomy Observatory, Box 443, Krugersdorp 1740, South Africa}
\address[robledo]{Centro de Astrobiolog\'{\i}a (INTA-CSIC), Ctra.~M-108, km.~4, E-28850 Torrej\'on de Ardoz, Madrid, Spain}
\address[f1]{Aalto University Department of Electronics and Nanoengineering, PL 15500, FI-00076 Aalto, Finland}
\address[f2]{Aalto University Mets\"ahovi Radio Observatory, Mets\"ahovintie 114, 02540 Kylm\"al\"a, Finland}
\address[s1]{Observatorio de Yebes (IGN), Cerro de la Palera SN, 19141, Yebes, Guadalajara, Spain}

\begin{abstract}
The largest Key Science Program of the \RA space VLBI mission is a survey of active galactic nuclei (AGN). 
The main goal of the survey is to measure and study the brightness of AGN cores in order to better understand the physics of their emission while taking interstellar scattering into consideration.
In this paper we present detection statistics for observations on ground-space baselines of a complete sample of radio-strong AGN at the wavelengths of 18, 6, and 1.3~cm. 
Two-thirds of them are indeed detected by \RA and are found to contain extremely compact, tens to hundreds of $\mu$as structures within their cores.
\end{abstract}
\end{frontmatter}

\section{Probing the emission mechanism of AGN jets}

The current paradigm for AGN assumes that their radio emission is synchrotron in nature and is produced by relativistic electrons. 
In this model, the intrinsic brightness temperatures cannot exceed $10^{11.5}$~K \citep{kel69,rea94}.
According to calculations by \citet{rea94}, it takes about a day for inverse Compton cooling (the so-called ``Compton Catastrophe'') to lower brightness temperatures initially exceeding such a limit due to, e.g., non-stationary injection of very high energy electrons in AGN cores, to values below this limit. However, the observed AGN emission might appear brighter due to Doppler boosting through bulk motion of the emitting plasma \citep[e.g.,][]{shkl64_jets}. Very long baseline interferometry (VLBI) kinematic studies of AGN show no evidence for Lorentz factors larger than 50 \citep{Cohen_etal07}, so Doppler boosting cannot increase the apparent jet brightness by more than a factor of about 100 over the intrinsic value. The typical boosting for blazar jets is expected on the level of about 10 or less \citep{MOJAVE_XIII}.
However, \textit{Fermi} gamma-ray and TeV Cherenkov telescope results introduce significant complications~--- the ``Doppler factor crisis.'' Compton models that explain these high energy parts of the spectrum including the very short timescale TeV flares \citep[e.g.,][]{2007ApJ...664L..71A,2007ApJ...669..862A}, require much larger Doppler factors than found from VLBI kinematics, and would imply observed radio core brightness temperatures higher than $10^{14}$\,K.

The highest brightness temperature that can be measured by a radio
interferometer does not depend on wavelength, but only on the physical
baseline length and the accuracy of the fringe visibility measurement
\citep[see, e.g.,][]{2cmPaperIV}. Thus, going to shorter
wavelengths does not help in measuring higher brightness temperatures.  The highest brightness temperatures measured for AGN from the ground are of the order of $10^{13}$~K
\citep[e.g.,][]{2cmPaperIV,lisakov17}.  
This finding is consistent with the earlier VLBI observations from space conducted during the TDRSS experiments \citep{1989ApJ...336.1098L,1989ApJ...336.1105L,1990ApJ...358..350L} and in the framework of the VLBI Space Observatory Programme \citep[VSOP,][]{fre02,VSOPsurvey4,VSOPsurvey5}. These observations probed baselines of up to 2.4 times the Earth diameter, but had a lower interferometric sensitivity compared to the more recent ground-based observations.
Further increasing the baseline length is the only practical way to measure much higher brightness temperatures, and hence, to address the Compton Catastrophe issue.  
\RA provides baselines up to 28 Earth diameters,
allowing measurements of brightness temperature up to
$10^{15}$--$10^{16}$~K. This capability offers an unprecedented
opportunity to place stringent observational constraints on the
physics of the most energetic relativistic outflows. 
We underline that prior to the \RA launch it was unknown if there were AGN compact and bright enough to be detected by a space VLBI system at baselines many times longer than the Earth diameter.
An indirect evidence that AGN contain regions of an angular diameter in the range of 10-50~$\mu$as was provided by IDV measurements \citep[e.g.,][]{lovell_etal08}.

\RA results on selected individual sources were presented earlier by 
\cite{Kovalev16,edwards17,pilipenko18,kutkin18} with an emphasis on the AGN brightness issue. 
In this paper we discuss \RA detection results for a complete VLBI-flux-density limited sample of bright AGN jets.

\section{Source sample and space VLBI observations}

\begin{figure}[tbh]
\centering
\includegraphics[width=0.6\textwidth,trim=0cm 0.5cm 0cm 0cm]{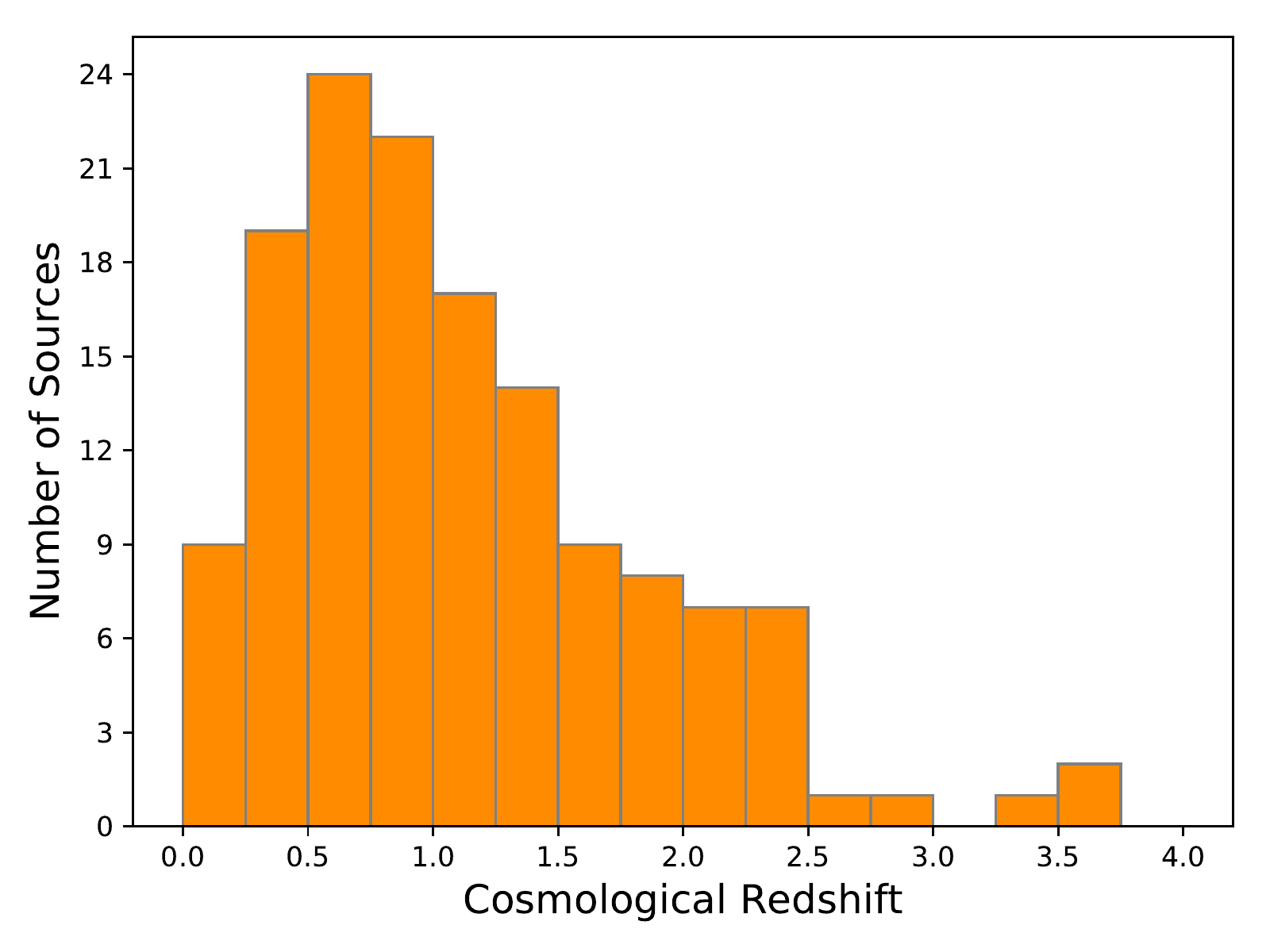}
\caption{Redshift distribution of the complete VLBI-flux-density limited sample of 163 compact extragalactic radio sources.
\label{f:z}
}
\end{figure}

The \RA AGN survey targets include the complete sample of 163 sources that have 8~GHz correlated flux densities at the ground baselines longer than 200~M$\lambda$ of $S_\mathrm{c}>600$~mJy as reported in the Radio Fundamental Catalog in 2012, at the time of the sample compilation\footnote{\url{http://astrogeo.org/rfc/}}. 
The large sample size is essential for modeling the complex selection
biases associated with relativistic beaming \cite[e.g.,][]{lister03}.
Fig.~\ref{f:z} presents the redshift distribution of these AGN.
The list of targets is augmented by AGN with jets showing the fastest speed \citep{MOJAVE_XIII}, strong scintillators selected from intraday variability (IDV) surveys \citep[e.g.,][]{lovell_etal08}, high redshift AGN, nearby AGN, and broad absorption line quasars. Here we discuss only the results related to the VLBI-flux-density limited sample. 

An overview of the \RA mission and the Spektr-R 10-m Space Radio Telescope (SRT) including its calibration is presented by \cite{ra1} and \cite{SRTrec2014}. The AGN Survey observations were performed independently at three observing bands: 1.3~cm (K), 6~cm (C), and 18~cm (L).
Terrestrially, the survey was supported by the following radio telescopes which have produced fringe detections with the SRT:
Arecibo 305\,m, 
phased Australia Telescope Compact Array (ATCA), 
Badary 32\,m, 
Ceduna 30\,m, 
Effelsberg 100\,m, 
Evpatoria 70\,m, 
Green Bank Telescope 100\,m, 
Hartebeesthoek 26\,m, 
Hobart 26\,m, 
Irbene 32\,m, 
Jodrell Bank 76\,m, 
Kalyazin 64\,m,
Medicina 32\,m, 
Mopra 22\,m, 
Noto 32\,m, 
Parkes 64\,m, 
Robledo 70\,m, 
Sheshan 25\,m,
Svetloe 32\,m, 
Tianma 65\,m,
Torun 32\,m, 
Usuda 64\,m, 
phased Karl G.~Jansky Very Large Array, 
phased Westerbork Synthesis Radio Telescope (WSRT), 
Yebes 40\,m, 
and Zelenchukskaya 32\,m.
The AGN survey was also supported by long-term monitoring of the broad-band total flux density at RATAN-600 (1.4-31~cm) and OVRO (2~cm) radio telescopes
as well as intra-day variability measurements by Effelsberg \citep{liu18}, ATCA, WSRT, and Urumqi.
The SRT recording rate was 128~Mbps with 1-bit sampling while ground telescopes utilized the 2-bit sampling with the total rate of 256~Mbps. The telescopes were recording $2\times 16$~MHz channels per polarization.

\RA detection sensitivity depends on the sensitivity of ground telescopes as well as coherence time for which we can integrate the data without significant losses. 
Accordingly, typical integration time at 18 and 6~cm was chosen to be up to 20~min while for 1.3~cm we have used 10~min long scans. Resulting detection sensitivity at the level of about $7\sigma$ with the largest ground telescopes was up to 6~mJy at 18 and 6~cm and 60~mJy at 1.3~cm.

The survey observations began within the \RA Early Science Program and have continued as one of the Key Science Programs, spanning the years May~2012 -- June~2016, inclusive.
Each single-source space VLBI observation lasted for 40-60 minutes and was split into scans that are 10-20 minutes long being supported on the ground typically by several telescopes per frequency band.
As the VLBI data collected by the SRT have to be downlinked to the ground in real time, a tracking station should be visible to the satellite's steerable high-gain antenna during the observations. This, together with the SRT Sun-avoidance angles and the ground telescopes' source visibility and scheduling constraints determine the planning of the survey observations. 
We used the \texttt{FakeRaT} software \citep{2015CosRe..53..216Z} based on the \texttt{FakeSat} code \citep{1991ASPC...19..107M,1994vtpp.conf...34M,2000AdSpR..26..637S} to model the SRT-related constraints and \texttt{SCHED}\footnote{\url{http://www.aoc.nrao.edu/software/sched/}} to compute source visibility and generate \texttt{vex} control files for the ground telescopes.
The Pushchino tracking station was utilized from the very beginning of the survey \citep{ra1}, while the Green Bank tracking station \citep{Ford14} joined the mission in August 2013.
The \RA VLBI experiments had to be separated by typically three-hour-long gaps to allow for the high-gain antenna drive to cool.
Given the above constraints, an effort was made to observe each source multiple times to cover the full range of accessible space-ground baselines.
The fast-evolving \RA orbit provided a different range of baselines and baseline position angles for a given source over the years during which the survey was conducted.
About 10\% of the complete sample had not been observed by June~2016. These are the low-declination targets, which are more difficult to schedule due to the limited availability of telescopes in the Southern Hemisphere and stronger visibility restrictions of the SRT due to the absence of a tracking station in the South.

The survey focused on total intensity measurements. 
To increase the outcome of the observations, the following observing scheme was chosen.
The SRT observed in a single-polarization dual-band mode. Typically, it was a combination of either L- and C-bands or C- and K-bands. 
An important advantage of this observing mode is the possibility of using the fringe detection from the lower frequency to check, or correct for, the Spektr-R orbit reconstruction uncertainty resulting in a large residual delay and its first and second derivatives for the higher frequency correlation and fringe search.

\begin{figure}[t]
\begin{centering}
\includegraphics[width=0.5\textwidth,trim=0cm 0cm 0cm 0cm]{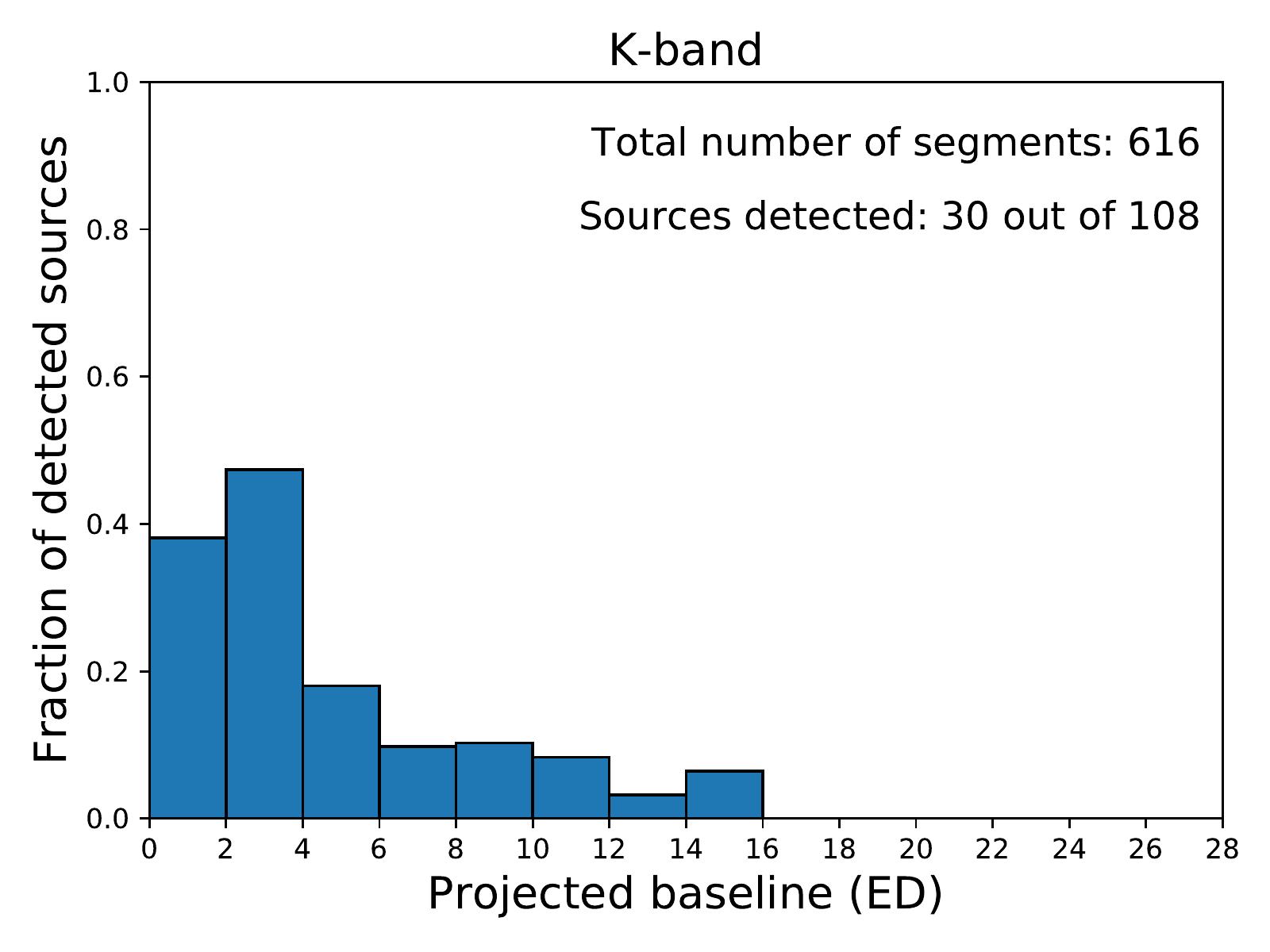}\\
\includegraphics[width=0.5\textwidth,trim=0cm 1.5cm 0cm 0cm]{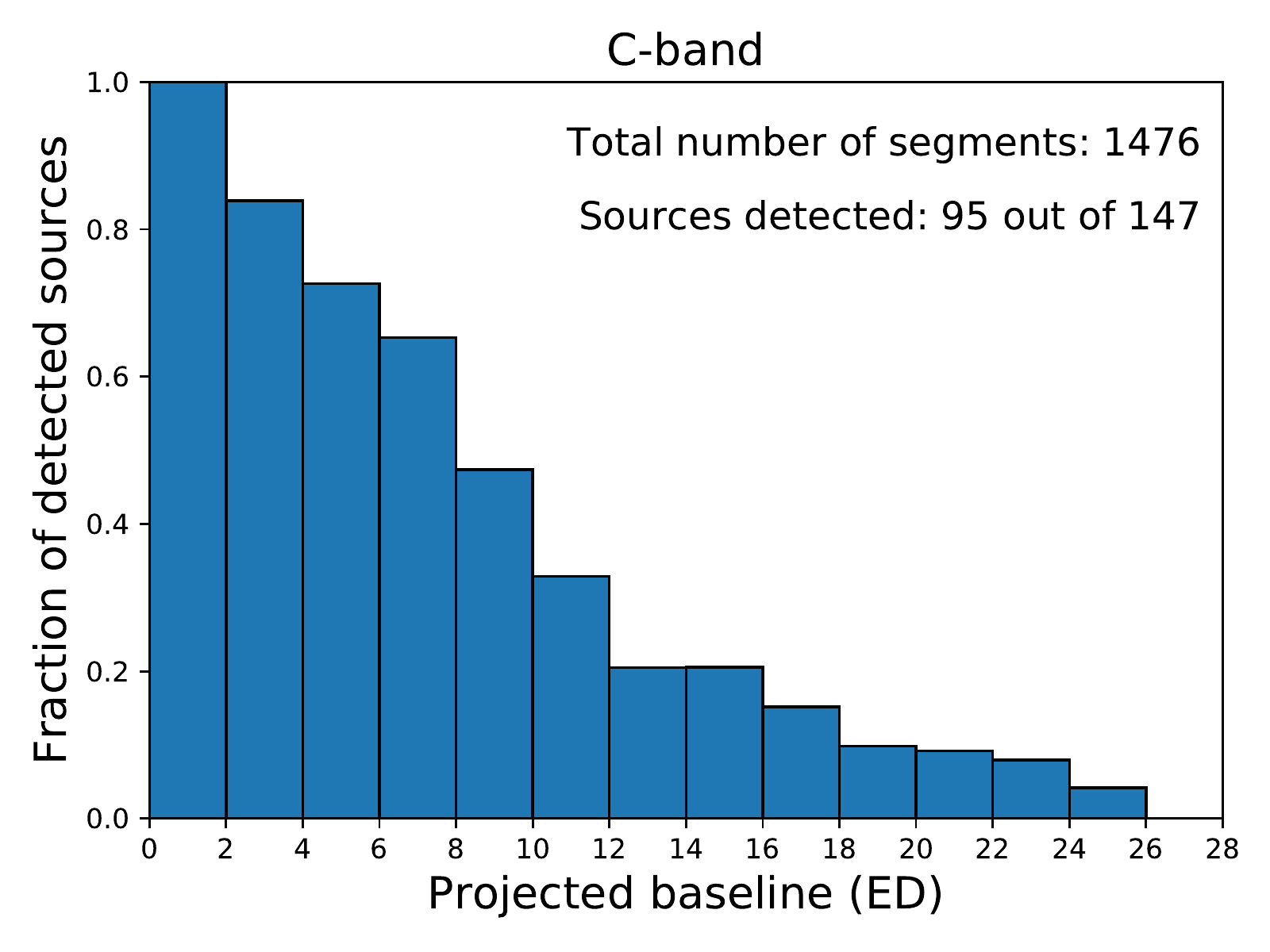}
\includegraphics[width=0.5\textwidth,trim=0cm 1.5cm 0cm 0cm]{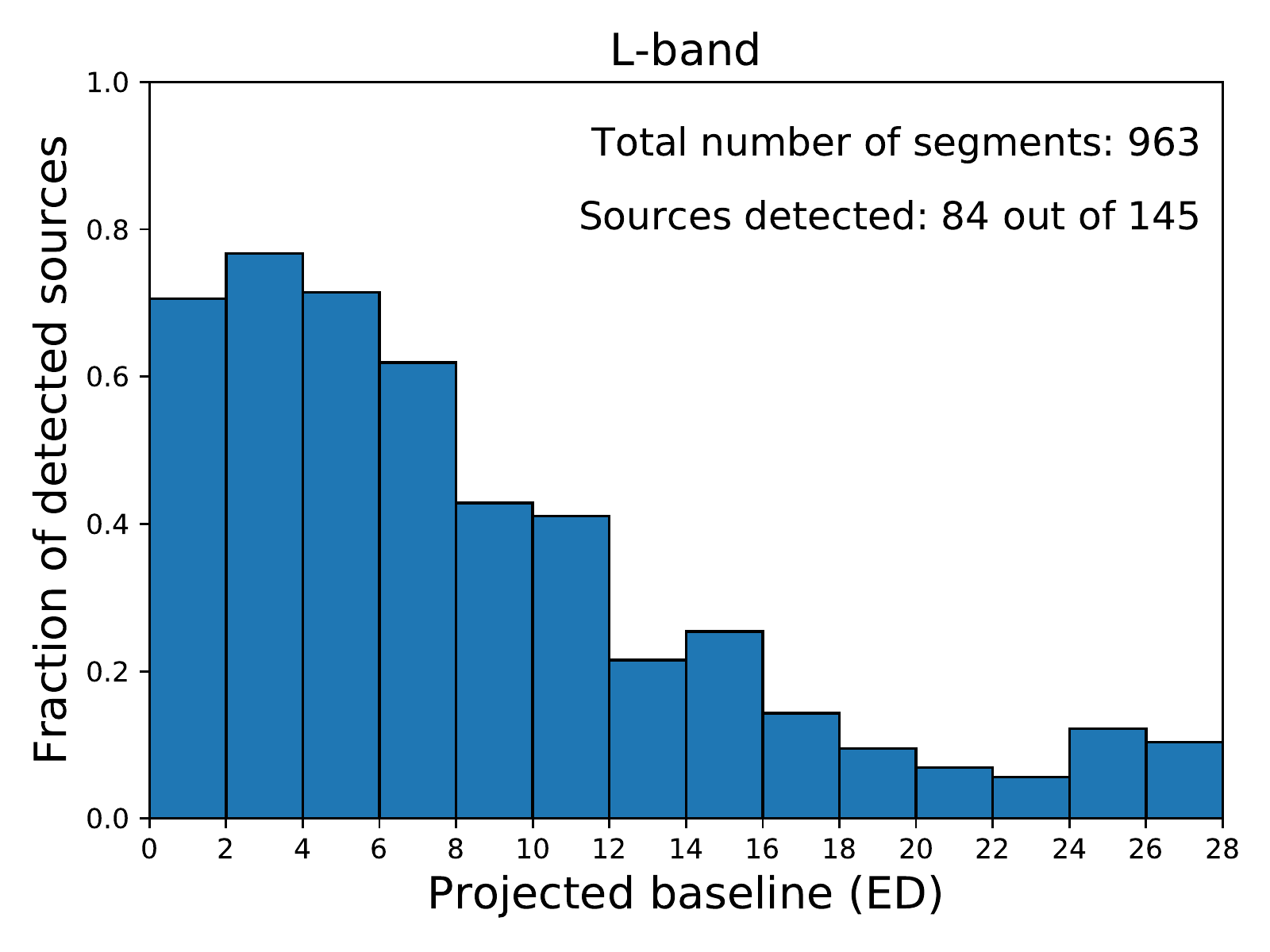}
\end{centering}
\caption{
Fraction of detected sources versus projected ground-space VLBI baselines (in units of Earth diameters, ED) for 1.3 (K-band), 6 (C-band), and 18~cm (L-band) observations.
\label{f:det}
}
\end{figure}

\begin{figure}[t]
    \centering
    \includegraphics[width=0.32\textwidth,trim=1.5cm 0cm 0cm 1cm,clip]{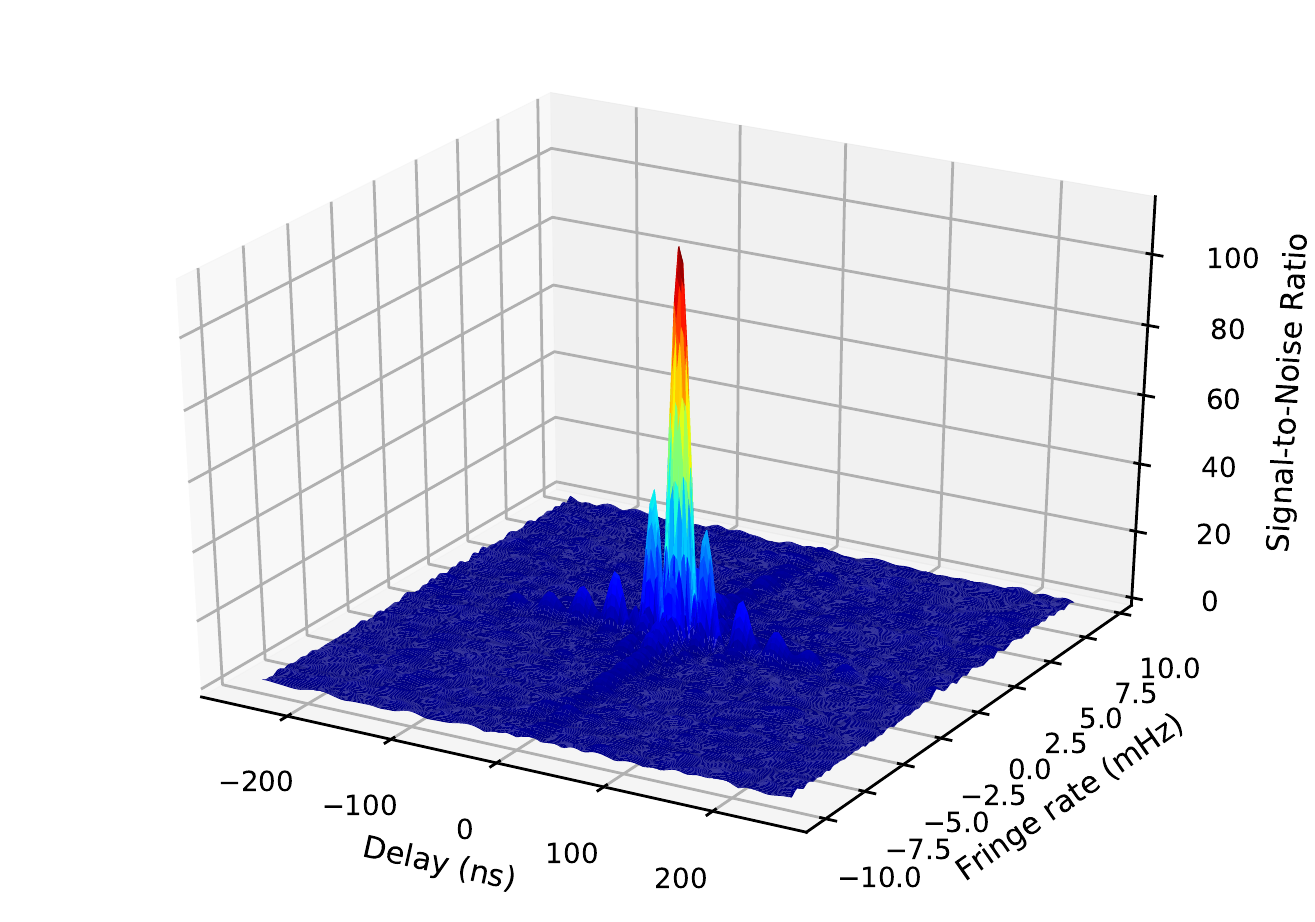}
    \includegraphics[width=0.32\textwidth,trim=1.5cm 0cm 0cm 1cm,clip]{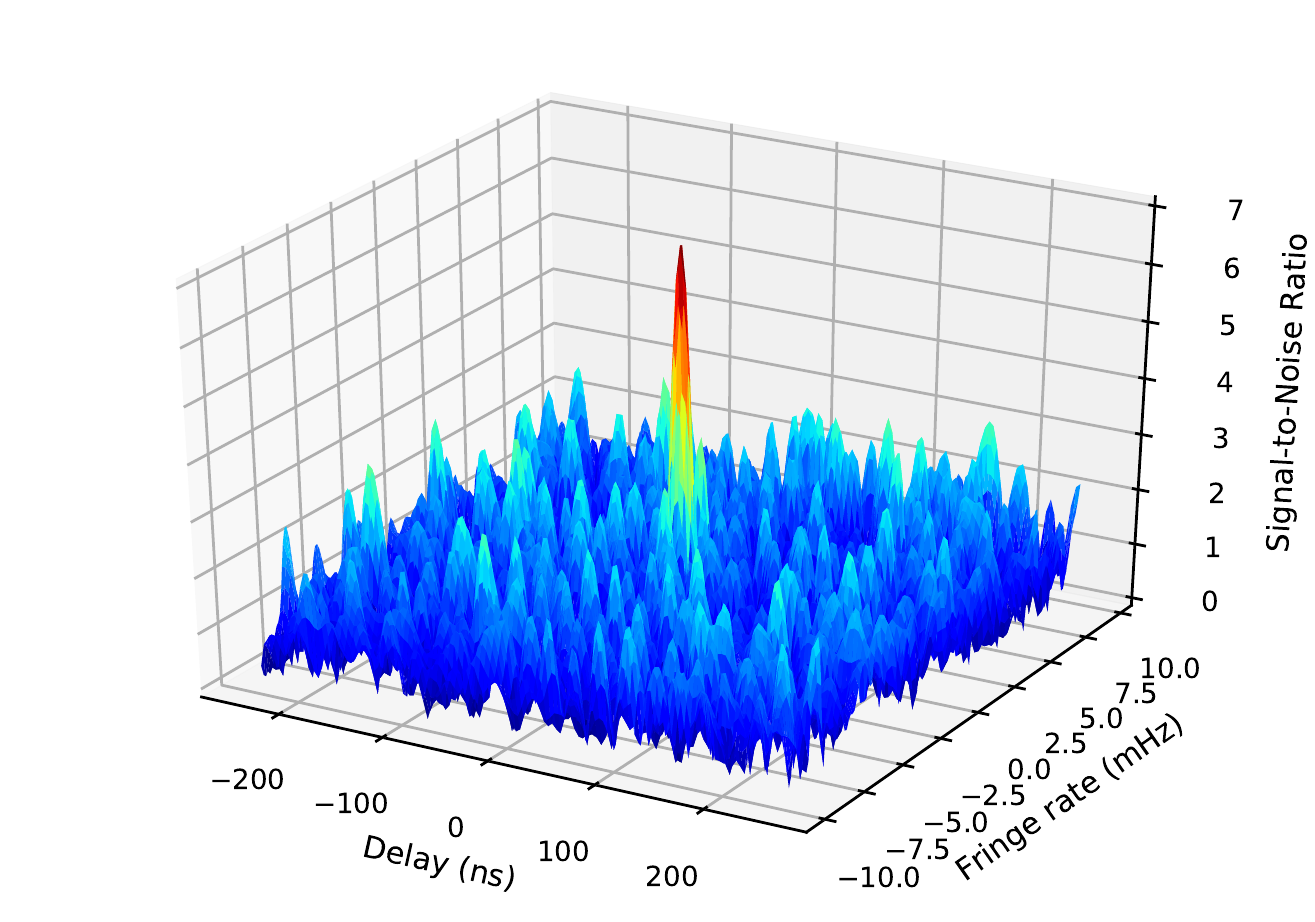}
    \includegraphics[width=0.32\textwidth,trim=1.5cm 0cm 0cm 1cm,clip]{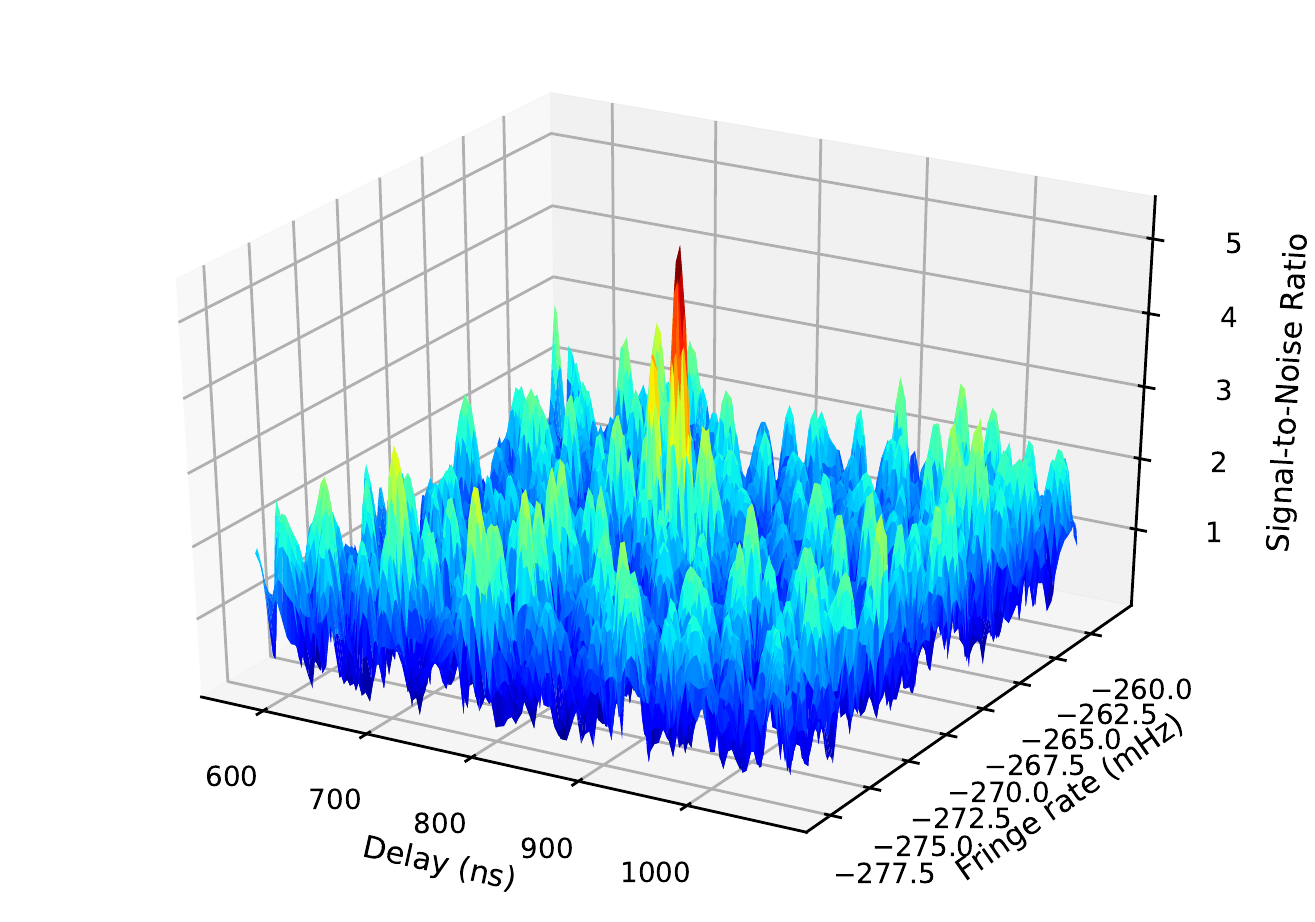}
    \caption{Examples of \RA interferometric fringes, C-band observations of 0716+714.
    \textit{Left panel:}
    Detection with $\textrm{PFD} < 10^{-100}$ at projected baseline 7.0~ED, SRT--Effelsberg, SNR=138;
    \textit{middle panel:}
    detection with $\textrm{PFD} = 10^{-8}$ at projected baseline 21.9~ED, SRT--Effelsberg, SNR=7.3;
    \textit{Right panel:}
    non-detection with $\textrm{PFD} = 0.07$ at projected baseline 22.0~ED, SRT--Noto, SNR=5.3
    \label{f:fringes}
    }
\end{figure}

\begin{figure}[t]
    \centering
    \includegraphics[width=0.6\textwidth,trim=0cm 0.5cm 0cm 0cm]{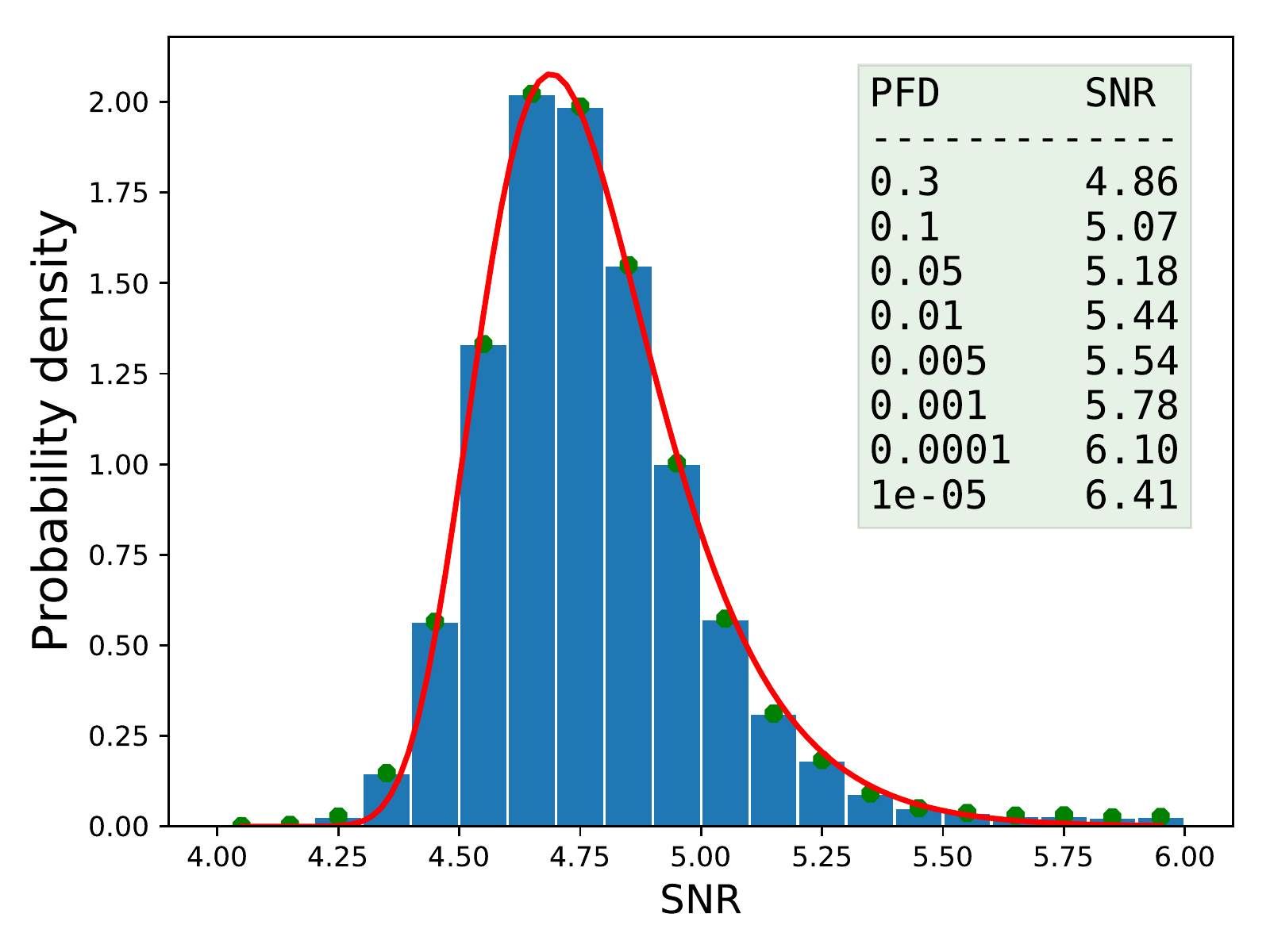}
    \caption{Low SNR part of the empirical distribution for the fringe SNR from the results of fringe fitting \RA AGN Survey data. This particular distribution is obtained from 6~cm data correlated with the following parameters: 64~spectral channels per 16-MHz wide frequency channel, correlator integration time 0.5~s, and 10~min fringe search interval.
    The red curve is the theoretical distribution \citep{VLBIbook} fitted to the low-SNR peak (the ``no signal'' case).
    The inset presents the correspondence between PFD and SNR for the given set of data parameters.
    \label{f:pfd}
    }
\end{figure}

\section{Space VLBI data analysis and detection results}

The data were correlated by the Astro Space Center \RA correlator in Moscow \citep{RAcorr} and post-processed with the \texttt{PIMA}\footnote{\url{http://astrogeo.org/pima/}} software \citep{VGaPS}.
The distribution of the fraction of detected sources versus projected \RA baseline is presented in Fig.~\ref{f:det} for the three observing bands separately. A detection is considered significant if the probability of a false detection (PFD) is less than 0.01\,\%. 
To determine the correspondence between the derived signal-to-noise ratio (SNR) and PFD for every observing scan we utilize the approach suggested by \citet{VGaPS}. 
We perform fringe fitting of AGN survey data and calculate an SNR statistic based on that (see examples in Fig.~\ref{f:fringes}). The low-SNR part of this distribution represents non-detected sources, and therefore fitting to this part of the distribution with a theoretical function allows us to relate observed SNR to probability of false detection (Fig.~\ref{f:pfd}). 
Note that the Figure presents the low-SNR part of the full set of SNR values only.
We determine the parameters of this probability density distribution for the used sets of the following parameters: the number of spectral channels in a 16-MHz frequency channel, correlator integration time, and scan lengths (i.e.\ fringe search interval). From these parameters we calculate the PFD value corresponding to a given value of the SNR for each observing scan. An example of the empirical SNR distribution and the theoretical probability density distribution fit is shown in Fig.~\ref{f:pfd}.
\RA has delivered detections in just over one third of the observing segments. 

The survey observations were scheduled at a low priority level for short projected spacings in order to allow AGN imaging, as well as pulsar, maser, and gravitational redshift observations to reach their goals.
As a result, the apparent drop in the detection fraction at 0 to 2 Earth Diameters (ED) can be observed in the K- and L-band histograms.
Moreover, the ground support of those survey observations was poorer than average. 
This scheduling issue explains the apparent drop of the detection fraction at short baselines, which should be treated as  an observational bias. The stronger statistics for the C-band observations results in a better, unbiased, first bin (Fig.~\ref{f:det}).

About two thirds of the observed complete sample are detected on space VLBI baselines. This means that many AGN jets, most probably their cores \citep{2cmPaperIV}, contain extremely compact regions of very bright synchrotron emission.
The AGN which are detected at extreme projected spacing about or longer than 25~ED at L-band include 0048$-$097, 0106+013, 0119+115, 0235+164, 0716+714, 1253$-$055 and at C-band 0235+164, 1124$-$186. At K-band, detections at baseline projections about or longer than 15~ED or 14~G$\lambda$ are found from 0235+164, 0716+714, 0851+202.
Many AGN detected by \RA are found to show brightness temperature values significantly in excess of the Compton Catastrophe limit and most of them are far above the equipartition value \citep{kel69,rea94}.

It is of interest to note that the fractional detection histograms look similar to the median normalized projected fringe plots generated by the 6\,cm VSOP \citep{VSOPsurvey4} and 2\,cm VLBA \citep{2cmPaperIV} surveys.
This basically reflects the core-jet structure of the observed targets but at the smaller scales probed.
To first order, the difference between the detection histograms can be attributed to the different sensitivities.
While the C-band and L-band observations have a comparable level of sensitivities, the
K-band data are significantly less sensitive due to the following three reasons: the efficiencies of both ground and space telescopes are lower; their system temperatures are higher; and the coherence time at 1.3~cm is significantly shorter than at 6 and 18~cm due to the Earth's troposphere.

We note a possible excess of fractional detections at the longest \RA projected baselines at 18\,cm in comparison to the 6\,cm results of about the same sensitivity. This can be an indication of the scattering sub-structure originally discovered in \RA pulsar observations \citep{Gwinn16,Popov17} and later confirmed by the ground-based observations of Sgr~A* \citep{gwinn_etal14,MJ18}. See also the analyses of \RA data for the quasars 3C\,273 \citep{MJ2016} and B\,0529+483 \citep{pilipenko18}.

These results also indicate that interstellar scattering only weakly affects the \RA 6~cm results and is completely absent in 1.3~cm data, for the typical mid-Galactic latitude sight-lines probed by the survey, following estimations by \citet{JG15}.
Full results and analysis of the \RA AGN survey data, as well as the methodology of \RA observations, are currently being finalized in a number of papers.

\section{Summary}

In this paper we have presented the results of detection statistics for a complete sample of 163 AGN jets from 18, 6, and 1.3~cm observations by the ground--space interferometer \textit{RadioAstron}.
Two thirds of the targets have delivered significant interferometric fringes at space VLBI baselines indicating the presence of ultra-compact and bright structures within cores in many of them.
An excess of 18~cm detections at the longest \RA baselines is attributed to the scattering sub-structure effect.

\section*{Acknowledgements}

We thank Ken~Kellermann, Chris~Salter, Kristen~Jones as well as anonymous referees for useful comments on the manuscript.
The \textit{RadioAstron} project is led by the Astro Space Center of the Lebedev Physical Institute of the Russian Academy of Sciences and the Lavochkin Scientific and Production Association under a contract with the State Space Corporation ROSCOSMOS, in collaboration with partner organizations in Russia and other countries.
The results are partly based on observations performed with radio telescopes of IAA RAS, the 100-m telescope of the MPIfR (Max-Planck-Institute for Radio Astronomy) at Effelsberg, the Medicina and Noto telescopes operated by INAF -- Istituto di Radioastronomia, 
the Sheshan and Tianma telescopes operated by Shanghai Astronomical Observatory of Chinese Academy of Sciences, the DSS-63 antenna at the Madrid Deep Space Communication Complex under the Host Country Radio Astronomy program.
The paper has used the Evpatoria RT-70 radio telescope (Ukraine) observations carried out by the Institute of Radio Astronomy of the National Academy of Sciences of Ukraine under a contract with the State Space Agency of Ukraine and by the National Space Facilities Control and Test Center with technical support by Astro Space Center of Lebedev Physical Institute, Russian Academy of Sciences.
This work is based in part on observations carried out using the 32-meter radio telescope operated by Torun Centre for Astronomy of Nicolaus Copernicus University in Torun (Poland) and supported by the Polish Ministry of Science and Higher Education SpUB grant.
The Hartebeesthoek telescope is a facility of the National Research Foundation of South Africa.
The Arecibo Observatory is operated by SRI International under a cooperative agreement with the National Science Foundation (AST-1100968), and in alliance with Ana G. Mendez-Universidad Metropolitana, and the Universities Space Research Association.
The Green Bank Observatory is a facility of the National Science Foundation operated under cooperative agreement by Associated Universities, Inc.
The Long Baseline Array is part of the Australia Telescope National Facility which is funded by the 
Australian Government for operation as a National Facility managed by CSIRO.
Results of optical positioning measurements of the Spektr-R spacecraft by the global MASTER Robotic Net \citep{MASTER2010}, ISON collaboration, and Kourovka observatory were used for spacecraft orbit determination in addition to mission facilities.


\end{document}